\documentstyle[psfig,prd,aps]{revtex}

\begin{document}
\draft

\title{Solution to the inverse problem for a noisy spherical gravitational 
wave antenna}

\author{Stephen M. Merkowitz}

\address{INFN Laboratori Nazionali di Frascati, Via Enrico Fermi 40, 
I-00044 Frascati (Roma) ITALY}

\date{June 10, 1998}

\maketitle

\begin{abstract}
A spherical gravitational wave antenna is distinct from other types of 
gravitational wave antennas in that only a single detector is necessary to 
determine the direction and polarization of a gravitational wave.  Zhou and 
Michelson showed that the inverse problem can be solved using the maximum 
likelihood method if the detector outputs are independent and have normally 
distributed noise with the same variance.  This paper presents an analytic 
solution using only linear algebra that is found to produce identical 
results as the maximum likelihood method but with less computational 
burden.  Applications of this solution to gravitational waves in 
alternative symmetric metric theories of gravity and impulsive excitations 
also are discussed.
\end{abstract}


\section{Introduction}

Several resonant-mass gravitational wave antennas are now in continuous 
operation with strain sensitivities of the order $10^{-21} 
\text{Hz}^{-1/2}$ \cite{Amaldi_1998}.  With further improvements to these 
detectors and the addition of several large laser interferometers now under 
construction \cite{VIRGO_1997}, the prospects for gravitational wave 
astronomy are quite good.  The underlying non-gravitational physics 
associated with these detectors is reasonably understood and further 
improvements can be based on solid technological guidelines.

Many believe the next generation of resonant-mass antennas will be of 
spherical shape \cite{Omni_1997}.  Confirmed detection of gravitational 
waves will require a coincidence between several detectors, thus the unique 
features of a sphere may play an essential role in a network of 
gravitational wave antennas.  Two important features of a sphere are its 
equal sensitivity to gravitational waves from all directions and 
polarizations and its ability to determine the directional information and 
tensorial character of a gravitational wave \cite{Wagoner_Pavia_1976}.

To take full advantage of these capabilities, one needs to be able to 
interpret the data such a detector will produce.  Recently, much work has 
been done to understand the output of a spherical antenna equipped with 
resonant transducers \cite{Johnson_PRL_1993,Lobo_EPL_1996}.  All of these 
proposals operate on the principle that the response of the transducers can 
be transformed into a quantity that has a one-to-one correspondence with 
the tensorial components of a gravitational wave.  With the measurement of 
these components, it is possible to solve the inverse problem to obtain the 
direction and polarization amplitudes of a gravitational wave.

The solution to the inverse problem for a noiseless spherical antenna first 
was outlined in the mid 1970's by Wagoner and 
Paik \cite{Wagoner_Pavia_1976}.  More recently, the solution for a network 
of five noiseless bar antennas or interferometers was solved by Dhurandhar 
and Tinto \cite{Dhurandhar_MNRAS_1988}.  This method assumed that the 
detectors were co-located but oriented in different directions.  This 
solution is quite elegant because the exact solution can be found using 
straightforward algebra.  Since a sphere can be thought of as five bar 
detectors occupying the same space, this solution can be adapted for a 
spherical antenna \cite{Zhou_PRD_1995,Magalhaes_RAS_1995}.  In addition, Lobo 
outlined a procedure that can be used if the correct theory of gravity is 
not general relativity but unknown \cite{Lobo_PRD_1995}.  What all these 
proposals have in common is that they use basic symmetry properties of the 
matrices describing the detectors and their response to a gravitational 
wave.  This makes them intuitive and easy to 
visualize \cite{Magalhaes_APJ_1997}.

The solution to the inverse problem in the presence of noise is more 
complicated.  G\"{u}rsel and Tinto solved the problem for three noisy 
interferometers using a maximum likelihood method \cite{Gursel_PRD_1989}.  
This solution required the measurement of the time delay of the signal 
between widely separated detectors to triangulate the direction of the 
source.  Zhou and Michelson showed that the inverse problem for a spherical 
antenna can be solved in the presence of noise, also using a maximum 
likelihood method \cite{Zhou_PRD_1995}.

What is disappointing about the maximum likelihood method is that the 
original simplicity of Dhurandhar and Tinto's noiseless solution is lost.  
In addition, an exact solution for the spherical detector was not found, 
making it necessary to solve the problem numerically.  This solution can be 
computationally expensive, especially if the signal-to-noise ratio (SNR) is 
low.  For the three interferometer case, the numerical solution often can 
lead to an incorrect estimate at low SNR \cite{Gursel_PRD_1989}.  This 
appears not to be a problem for a spherical antenna; a global maximum 
usually exists and is aligned with the correct direction.  This leads us to 
believe that the problem for a spherical antenna can be solved 
analytically.

On experiments with the room-temperature prototype spherical antenna (TIGA) 
at Louisiana State University, we used a procedure to solve the inverse 
problem for impulsive excitations applied to the sphere surface 
\cite{Merkowitz_PRD_1996,Merkowitz_PRD_1997}.  This solution was similar to 
Dhurandhar and Tinto's original solution for gravitational waves, but we 
used a perturbation argument (presented below) to take into account the 
finite SNR of the experiment.  The case of an impulsive excitation is more 
simple than a gravity wave because fewer parameters are involved, however, 
we show below that this method also can be used to find an analytic 
solution for gravitational waves.

We begin by reviewing the response of a spherical antenna to gravitational 
waves in general relativity and show how Dhurandhar and Tinto's original 
method can be applied to solve for the wave direction and polarization 
amplitudes.  We then generalize the arguments to any symmetric metric 
theory of gravity as well as to impulsive excitations.  In 
Sec.~\ref{sec:noisy} we show how this technique can be extended to a noisy 
antenna with independent and equally sensitive detector outputs.  This 
solution is found to be equivalent to the maximum likelihood method under 
the same noise requirements.  The general approach taken allow this 
solution to be easily adapted to other types of excitations with similar 
symmetry properties.  We conclude the paper with a discussion of the 
limitations of the solution and possible extensions of this method.

\section{Detector response of an elastic sphere}
\label{sec:interaction}

Dhurandhar and Tinto solved the inverse problem for 5 bar antennas as well 
as 5 interferometers \cite{Dhurandhar_MNRAS_1988}.  Others have used their 
method to solve the problem for a spherical antenna 
\cite{Zhou_PRD_1995,Magalhaes_RAS_1995}.  Their technique involves 
constructing a matrix, say $\bbox{A}$, that describes the response of the 
detector to a gravitational wave.  They found that in general relativity 
the eigenvector of $\bbox{A}$ with zero eigenvalue points in the 
propagation direction of the wave.  In the following we will also use this 
concept, but will derive the equations in the context of linear algebra as 
this will lead us directly into the solution for the noisy antenna.  For a 
more complete discussion of the response of an elastic sphere to 
gravitational waves the reader is referred to 
Refs.~\cite{Ashby_PRD_1975,Merkowitz_PRD_1995,Lobo_PRD_1995}.

\subsection{Detector response in general relativity}

A gravitational wave is a traveling time-dependent deviation of the metric 
perturbation, denoted by $H_{\mu\nu}(t)$.  We follow a common textbook 
development for the metric deviation of a gravitational wave, which finds 
that only the spatial components $H_{ij}(t)$ are non-zero, and further can 
be taken to be transverse and traceless \cite{Thorne_300y}.  This tensor is 
simplified if we initially write it in the ``wave-frame,'' denoted by 
primed coordinates and indices.  This is a coordinate frame with origin at 
the center of mass of the detector and the $z'$ axis aligned to the 
propagation direction of the wave.  We restrict ourselves to detectors much 
smaller than the gravitational wavelength so only the time dependence of 
$H_{ij}(t)$ will have significant physical effects.  A general form for the 
spatial components of the metric deviation in the wave-frame can be written 
as
\begin{equation}
   \bbox{H}'(t) = \left[ 
   \begin{array}{ccc}
      h_+(t)      & h_\times(t) & 0 \\
       h_\times(t) &  -h_+(t)    & 0 \\
      0            & 0            & 0
   \end{array} \right],
   \label{eqn:wave_strain}
\end{equation}
where $h_+(t)$ and $h_\times(t)$ are the wave amplitudes for the two 
allowed states of linear polarization and are called the plus and cross 
amplitudes.

The detector is more easily described in the ``lab-frame,'' denoted by 
unprimed coordinates and indices, with origin at the center of mass of the 
detector and the $z$ axis aligned with the local vertical.  In this frame, 
the primary physical effect of a passing gravitational wave is to produce a 
time dependent ``tidal'' force density $f^{\text{GW}}(\bbox{x},t)$ on the 
material with mass density $\rho$ at coordinate location $x_i$.  This force 
is related to the metric perturbation by
\begin{equation}
   f^{\text{GW}}_i(\bbox{x},t) = \frac{1}{2} \rho \sum_{j} 
   \ddot{H}_{ij}(t) x_j.
   \label{accel_metric_pert}
\end{equation}

It is natural to look for an alternate expression that separates the 
coordinate dependence into radial and angular parts.  Because the tensor 
$H_{ij}(t)$ is traceless, the angular expansion can be done completely with 
the five second order real valued spherical harmonics 
$Y_{2m}(\theta,\phi)$, where the index $m = 1\ldots5$.  We call the 
resulting time dependent expansion coefficients, denoted by $h_m(t)$, the 
``spherical amplitudes'' \cite{Merkowitz_PRD_1995}.  They are a complete 
and orthogonal representation of the cartesian metric deviation tensor 
$H_{ij}(t)$.  They depend only on the two wave-frame amplitudes and the 
direction of propagation.

\begin{figure}
\centerline{\psfig{file=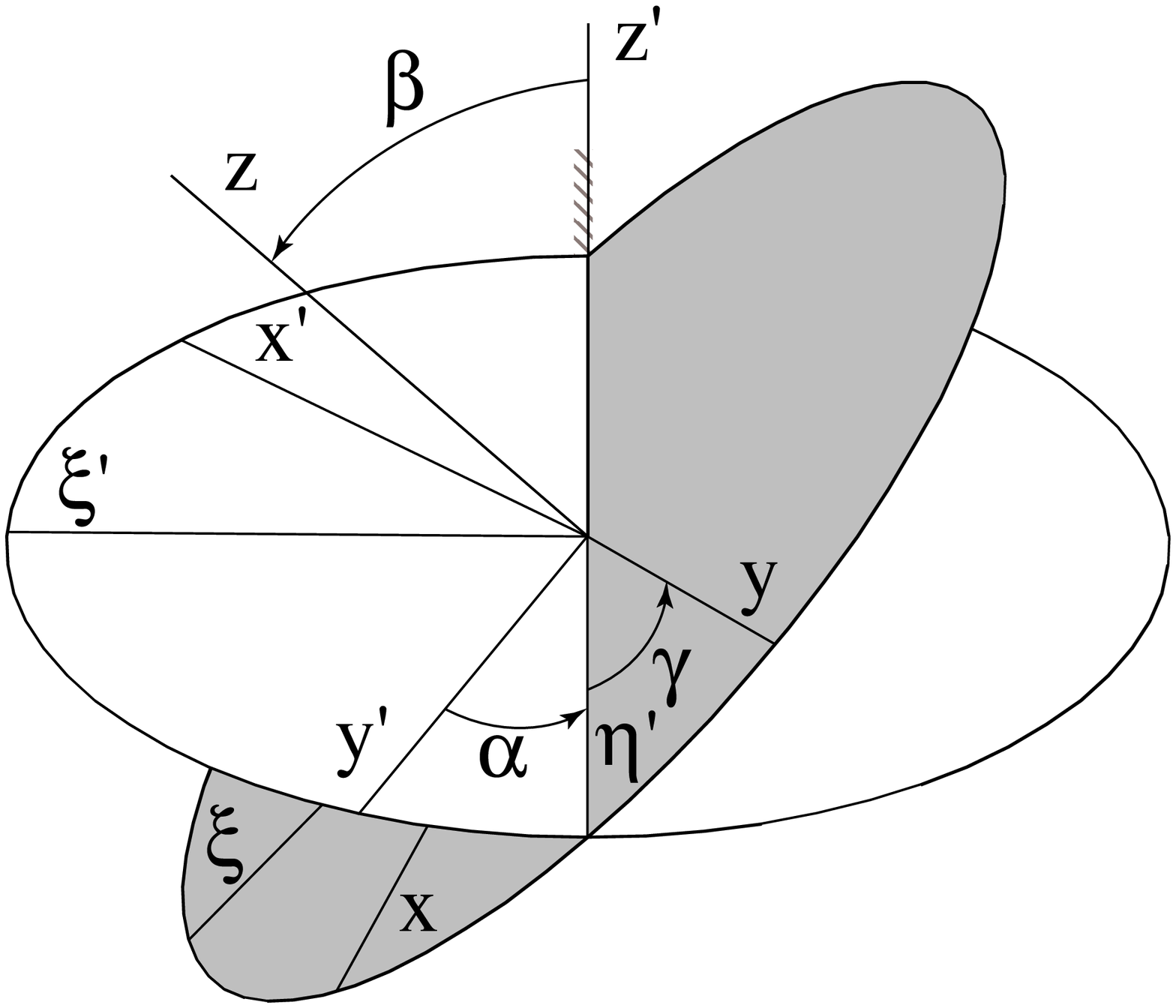,width=3.0in}}
\caption{The y-convention of the Euler angles.}
\label{fig:y-convention}
\end{figure}

To transform the metric perturbation to the lab-frame we perform the 
appropriate rotations using the y-convention of the Euler angles shown in 
Fig.~\ref{fig:y-convention}.  We denote the rotation about the wave $z'$ 
axis by $\alpha$, the rotation about the new $y$ axis ($\eta'$) by $\beta$, 
and the rotation about the final lab $z$ axis by $\gamma$.  The rotation 
matrix for the y-convention is
\begin{equation}
	\bbox{R} = 
	\left[ \begin{array}{ccc}
		\cos{\gamma}\cos{\beta}\cos{\alpha}-\sin{\gamma}\sin{\alpha} &
		\cos{\gamma}\cos{\beta}\sin{\alpha}+\sin{\gamma}\cos{\alpha} & 
		-\cos{\gamma}\sin{\beta} 
		\\
		-\sin{\gamma}\cos{\beta}\cos{\alpha}-\cos{\gamma}\sin{\alpha} & 
		-\sin{\gamma}\cos{\beta}\sin{\alpha}+\cos{\gamma}\cos{\alpha} &
		\sin{\gamma}\sin{\beta}  
		\\
		\sin{\beta}\cos{\alpha} &
		\sin{\beta}\sin{\alpha} &
		\cos{\beta}
	\end{array} \right].
	\label{eqn:rotation_matrix_full}
\end{equation}
At this point we arbitrarily set the rotation $\alpha$ about the wave $z'$ 
axis equal to zero; inclusion of this rotation will only ``mix'' the two 
polarizations of the wave.  The spherical amplitudes can now be written in 
terms of the polarization amplitudes and the source direction
\begin{mathletters}
\label{eqn:h_all}
\begin{eqnarray}
   h_1(t) 
   & = & 
   h_+(t)\frac{1}{2}\left( {1+\cos ^2\beta } \right)\cos 
   2\gamma + h_\times(t) \cos \beta \sin 2\gamma,
   \label{eqn:h1} \\
   h_2(t) & = & 
   -h_+(t)\frac{1}{2}\left( {1+\cos ^2\beta } \right)\sin 2\gamma 
   + h_\times(t) \cos \beta \cos 2\gamma,
   \label{eqn:h2} \\
   h_3(t) 
   & = & 
   -h_+(t)\frac{1}{2}\sin 2\beta \sin \gamma + h_\times(t) 
   \sin\beta \cos\gamma,
   \label{eqn:h3} \\
   h_4(t) 
   & = & 
   h_+(t)\frac{1}{2}\sin 2\beta \cos \gamma + h_\times(t) 
   \sin\beta \sin\gamma,
   \label{eqn:h4} \\
   h_5(t) 
   & = & 
   h_+(t)\frac{1}{2}\sqrt 3\sin ^2\beta.
   \label{eqn:h5}
\end{eqnarray}
\end{mathletters}

The mechanics of a spherical antenna can be described by ordinary elastic 
theory.  One finds that the eigenfunctions of an uncoupled sphere can be 
written in terms of the spherical harmonics
\begin{equation}
	\bbox{\Psi}_{n\ell m}(r,\theta,\phi) = \left({
	\alpha_{n\ell}(r)\bbox{\hat r} + \beta_{n\ell}(r) a \bbox{\nabla} 
	}\right) Y_{\ell m}(\theta,\phi).  
	\label{eqn:sphere_eigf}
\end{equation}
The radial eigenfunctions $\alpha_{n\ell}(r)$ and $\beta_{n\ell}(r)$ 
determine the motion in the radial and tangential directions respectively 
and depend on the radius $a$ and the material of the 
sphere \cite{Wagoner_Pavia_1976,Merkowitz_PRD_1995}.

In general relativity, only the 5 quadrupole modes of vibration will 
strongly couple to the force density of a gravitational wave.  For an ideal 
sphere they are all degenerate, having the same eigenfrequency, and are 
distinguished only by their angular dependence.  The effective force 
$F_{12m}(t)$ that a gravitational wave will exert on a fundamental 
quadrupole mode~$m$ of the sphere is given by the overlap integral between 
the eigenfunctions of the sphere and the gravitational tidal force
\begin{equation}
   F_{12m}(t) 
   \equiv 
   \int{\bbox{\Psi}_{12m}(\bbox{x}) \cdot 
   \bbox{f}^{\text{GW}}(\bbox{x},t)\,d^3x}
   = 
   \frac{1}{2} \ddot{h}_m(t) \, M \, \chi a.
\label{eqn:eff_force}
\end{equation}
Each spherical component of the gravitational field determines uniquely the 
effective force on the corresponding mode of the sphere and they are all 
identical in magnitude.  We can interpret the effective force $F_{12m}(t)$ 
in each mode as the product of: the physical mass of the sphere $M$, an 
effective length $\chi a$ (a fraction of the sphere radius), and the 
gravitational acceleration $\frac{1}{2}\ddot{h}_m(t)$.  The value of the 
coefficient $\chi$ depends on the sphere material, but is typically $\simeq 
0.6$ \cite{Merkowitz_PRD_1995}.

By monitoring the quadrupole modes of the sphere, one has a direct 
measurement of the effective force on the sphere and thus the spherical 
amplitudes of the gravitational wave.  The standard technique for doing so 
on resonant detectors is to position resonant transducers on the surface of 
the sphere that strongly couple to the quadrupole modes.  A number of 
proposals have been made for the type and positions of the transducers 
\cite{Johnson_PRL_1993,Lobo_EPL_1996,Zhou_PRD_1995}.  What all of these 
proposals have in common is that the outputs of the transducers are 
combined into ``mode channels'' $g_m(t)$ that are constructed to have a 
one-to-one correspondence with the quadrupole modes of the sphere and thus 
the spherical amplitudes of the gravitational wave 
\cite{Merkowitz_PRD_1995,Lobo_CQG_1998},
\begin{equation}
	g_m(t) \propto F_{12m}(t) \propto h_m(t).
\end{equation}
The mode channels can be collected to form a ``detector response'' matrix 
$\bbox{A}(t)$ that in the absence of noise is equal to the cartesian strain 
tensor $\bbox{H}(t)$ in the lab frame
\begin{equation}
   \bbox{A}(t) 
   \equiv
   \left[\begin{array}{ccc}
      g_{1}(t) - \frac{1}{\sqrt{3}}g_{5}(t) & g_{2}(t) &  g_{4}(t) \\
      g_{2}(t) &  -g_{1}(t) - \frac{1}{\sqrt{3}}g_{5}(t) &  g_{3}(t) \\
      g_{4}(t) &  g_{3}(t) &  \frac{2}{\sqrt{3}} g_{5}(t)
   \end{array} \right].
   \label{eqn:cartesian_strain_tensor}
\end{equation}  
For the remainder of this discussion we drop the notation of time 
dependence $(t)$ for brevity.

The strain tensor in the lab frame $\bbox{H}$ is a symmetric traceless 
matrix.  Consequently, it can be orthogonally diagonalized and has an 
orthonormal set of three eigenvectors.  One can construct from the 
eigenvectors a transformation matrix $\bbox{R}$ that diagonalizes 
$\bbox{H}$.  The matrix $\bbox{R}$ is also orthogonal, thus it can be 
considered a rotation matrix (it may also include a reflection).  The 
physical interpretation of this transformation is to rotate the lab frame 
such that the $z$ axis points in the direction of the source.  The matrix 
$\bbox{R}$ (the eigenvectors) will tell us the angles of rotation and thus 
the direction of the wave.

In the wave frame, $\bbox{H}'$ is not normally diagonal but it can be 
diagonalized by rotating Eq.~(\ref{eqn:wave_strain}) about the propagation 
axes using the Euler angle $\alpha$.  $\alpha$ may be a constant or a 
function of time depending upon the situation.  This rotation changes the 
polarization components of the tensor but not the wave direction relative 
to the lab frame.

To calculate the rotation matrix $\bbox{R}$ we need to solve the general 
eigenvalue equation for the strain tensor
\begin{equation}
	\bbox{H} \; \bbox{x} = \lambda_x \bbox{x}.
	\label{eqn:general_strain_ev}
\end{equation}
Since $\bbox{A}$ and $\bbox{H}$ are equal in the absence of noise we are 
free to substitute $\bbox{A}$ in Eq.~(\ref{eqn:general_strain_ev}) for 
$\bbox{H}$.  By inspection of Eq.~(\ref{eqn:wave_strain}) we see that in 
general relativity the eigenvector of $\bbox{H}$ with $\lambda_x = 0$ 
points in the propagation direction of the wave.  The direction can be 
calculated from this eigenvector by recognizing that it corresponds to the 
last column vector of $\bbox{R}$ in Eq.~(\ref{eqn:rotation_matrix_full}).  
Dividing the elements of this column we find
\begin{equation}
   \tan{\gamma} = -\frac{y}{x}, 
   \label{eqn:gamma}
\end{equation}
\begin{equation}
	\tan{\beta} = \frac{y}{z}\frac{1}{\sin{\gamma}}. 
	\label{eqn:beta}
\end{equation}	
The unusual minus sign in Eq.~(\ref{eqn:gamma}) comes from the use of the 
y-convention of the Euler angles.  Expanding 
Eq.~(\ref{eqn:general_strain_ev}) for $\lambda_x = 0$ and substituting in a 
particular choice of matrix elements from 
Eq.~(\ref{eqn:cartesian_strain_tensor}) we find
\begin{equation}
   \tan{\gamma}
   =
   \frac{3 g_{4}g_{3} - 2\sqrt{3}g_{2}g_{5}}
   {2\sqrt{3}g_{1}g_{5} + 2 g_{5}^2  + 3 g_{3}^2},
	\label{eqn:noisless_gamma}
\end{equation}
\begin{equation}
	\tan{\beta} 
	=
	\pm \frac{\sqrt{3}g_{3}g_{4} - 2 g_{5}g_{2}}
	{\sqrt{3}g_{1}g_{4} + g_{5}g_{4} + \sqrt{3}g_{3}g_{2}}
	\frac{1}{\sin{\gamma}}.
	\label{eqn:noisless_beta}
\end{equation}

This solution is valid only for a noiseless antenna; it will fail otherwise 
because we can no longer replace $\bbox{H}$ with $\bbox{A}$ and their 
eigenvectors and eigenvalues will no longer be equal.  The $\pm$ in 
Eq.~(\ref{eqn:noisless_beta}) illustrates the unavoidable fact that a 
single sphere cannot distinguish between antipodal sources.  This ambiguity 
is a characteristic of all gravity wave detectors, but can be removed by 
measuring the time delay of the signal between two widely separated 
antennas.

Once the direction is calculated we can determine the two polarization 
amplitudes by taking a linear combination of Eqs.~(\ref{eqn:h_all}).  These 
equations are actually overdetermined so several solutions exist (we have 5 
equations but only 4 unknowns).  In the absence of noise any particular 
solution to them is valid, but in anticipation of the noisy case we will 
take a systematic approach to the solution.

We need only the angles $\beta$ and $\gamma$ to rotate $\bbox{H}$ to 
$\bbox{H}'$, so at this point we again set $\alpha=0$.  The amplitudes are 
found by equating them to the corresponding matrix elements of $\bbox{H}'$ 
in Eq.~(\ref{eqn:wave_strain}).  Again, we may substitute $\bbox{A}$ for 
$\bbox{H}$ and $\bbox{A}'$ for $\bbox{H}'$ so we have $h_+ = A_{11}' = 
-A_{22}'$ and $h_\times = A_{12}' = A_{21}'$.  $\bbox{A}$ and $\bbox{A}'$ 
are symmetric so $A_{12}'$ and $A_{21}'$ will always be identical even when 
noise is introduced.  However, no such restriction is placed on $A_{11}'$ 
and $A_{22}'$.  We will use the average $(A_{11}'-A_{22}')/2$ to calculate 
$h_+$ for reasons that will become clear later.

Multiplying $\bbox{A}'=\bbox{R}^T \bbox{A} \bbox{R}$ for $\alpha=0$ and 
selecting the proper elements we find
\begin{equation}
	h_+ =
	  g_{1} \frac{1}{2}(1+\cos^{2}{\beta})\cos{2\gamma} 
	- g_{2} \frac{1}{2}(1+\cos^{2}{\beta})\sin{2\gamma}
	- g_{3} \frac{1}{2}\sin{2\beta}\sin{\gamma}
	+ g_{4} \frac{1}{2}\sin{2\beta}\cos{\gamma}
	+ g_{5} \frac{\sqrt{3}}{2}\sin^{2}{\beta}.
	\label{eqn:h_p}
\end{equation}
\begin{equation}
	h_\times =
	g_{1} \cos{\beta}\sin{2\gamma} + 
	g_{2} \cos{\beta}\cos{2\gamma} +
	g_{3} \sin{\beta}\cos{\gamma} + 
	g_{4} \sin{\beta}\sin{\gamma}.
	\label{eqn:h_x}
\end{equation}
These equations can also be derived by taking a linear combination of 
Eqs.~(\ref{eqn:h_all}).  This is not the only valid solution in the 
noiseless case, but it is particularly symmetric: the coefficients of each 
component $g_m$ is the same as the corresponding coefficients of $h_+$ or 
$h_\times$ in Eqs.~(\ref{eqn:h_all}) for $h_m$.  The fact that $h_\times$ 
does not contain a $g_5$ contribution is an artifact of using the 
y-convention of the Euler angles; in other conventions this term may be 
non-zero.

\subsection{Detector response in alternative theories of gravity}
\label{sec:theories}

Experiments in the solar system and pulsar-timing tests have ruled out many 
competing theories of gravity, however, general relativity is not the only 
theory of gravity that passes these weak field tests \cite{Will_TEGP_1993}.  
One measurement that can potentially rule out certain gravitational 
theories is the properties of gravitational waves \cite{Eardley_PRL_1973}, 
such as the speed of propagation and allowable polarization states.  It was 
shown above how a single sphere can measure the quadrupole components of 
the strain tensor, but a scalar wave can excite both the monopole mode and 
the quadrupole modes of a sphere \cite{Lobo_PRD_1995,Bianchi_PRD_1998}.  By 
monitoring both types of modes, a single spherical detector can measure all 
the tensor components of a gravitational wave.  This makes it possible for 
a single spherical detector to determine all of the six polarization states 
predicted by the most general symmetric metric theory of gravity 
\cite{Bianchi_CQG_1996}.

We can rewrite Eq.~(\ref{eqn:eff_force}) in terms of the electric 
components of the Riemann tensor \cite{Eardley_PRL_1973} $E_{ij}=R_{0i0j}$,
\begin{equation}
	F_{n\ell m} = 
	- \frac{1}{M} E_{ij} \int \Psi_{n\ell m}^{i} x^{j}\rho d^{3}x,
	\label{eq:electric_force}
\end{equation}
where we now include both the $\ell=2$ quadrupole modes and the $\ell=0$ 
monopole mode.  The monopole mode of an elastic sphere is actually at a 
higher frequency than the quadrupole modes.  If the source is not wide-band 
enough for detection in both of these modes, a second sphere with the 
monopole mode tuned to the quadrupole modes of the first will be needed to 
measure this component.  If the first sphere is at relatively low 
frequency, one might consider making the second sphere hollow to keep it of 
a practical size \cite{Coccia_PRD_1998}.  An alternative to a second sphere 
is to monitor the $n=2$ quadrupole modes and the monopole mode of a single 
sphere.  These modes are not far in frequency from each other and also have 
relatively large cross-sections \cite{Coccia_PRD_1995,Lobo_PRD_1995}.

Expanding Eq.~(\ref{eq:electric_force}) into radial and angular parts we 
find an additional spherical amplitude $h_{0}$ corresponding to the 
$\ell=0$ spherical harmonic.  The detector response in the lab frame can 
now be written as
\begin{equation}
   \bbox{A} 
   \equiv
   \left[\begin{array}{ccc}
	   g_1 - \frac{1}{\sqrt{3}}g_5 + g_0 & g_2 &  g_4 \\
	   g_2 & -g_1 - \frac{1}{\sqrt{3}}g_5 + g_0 & g_3 \\
	   g_4 &  g_3 & \frac{2}{\sqrt{3}}g_5 + g_0
   \end{array} \right].
   \label{eq:general_response}
\end{equation}

To determine how to solve the inverse problem we need to examine the form 
of $E_{ij}$.  It is a symmetric tensor so it has only six independent 
components.  It can be written in terms of the complex Newman-Penrose 
parameters \cite{Newman_JMP_1962} which allow the identification of the 
spin content of the metric theory responsible for the generation of the 
wave
\begin{equation}
	E_{ij} = \left[
	\begin{array}{ccc}
		-\Re\Psi_4-\Phi_{22} & \Im\Psi_4           & -\sqrt8 \Re\Psi_3 \\
		 \Im\Psi_4           & \Re\Psi_4-\Phi_{22} &  \sqrt8 \Im\Psi_3 \\
		-\sqrt8 \Re\Psi_3    & \sqrt8 \Im\Psi_3   & -6\Psi_2
	\end{array}\right].
\end{equation}

We can divide the theories of gravity into categories using the E(2) 
classification scheme shown in Tab.~\ref{tab:E2_class} 
\cite{Eardley_PRD_1973}.  The tensor $E_{ij}$ is symmetric for all of these 
classes, thus it is orthogonally diagonalizable, but classes $II_6$ and 
$III_5$ have more degrees of freedom (direction plus polarization states) 
than we are capable of measuring with a single spherical detector.  These 
two classes are often referred to as ``observer-dependent'' because 
different observers will disagree upon which polarization states are 
present.  As a consequence, the polarization amplitudes for a particular 
observer must be known before the direction of the wave can be estimated.

For the ``observer-independent'' classes $O_0$, $O_1$, $N_2$, and $N_3$, 
the situation is more straightforward.  $O_0$ is obviously uninteresting as 
it does not predict any gravitational waves (this class along with $O_1$ 
have essentially been ruled out by previous experiments 
\cite{Will_TEGP_1993}).  We notice that $E_{ij}$ for the 
observer-independent classes can be diagonalized by a rotation $\alpha$ 
about the propagation axis, therefore, we can use the same arguments 
presented above for general relativity to solve for the wave direction.

The most general observer-independent class is $N_3$, which has
\begin{equation}
	E_{ij} = \left[
	\begin{array}{ccc}
		-\Re\Psi_4-\Phi_{22} & \Im\Psi_4           & 0 \\
		 \Im\Psi_4           & \Re\Psi_4-\Phi_{22} & 0 \\
		 0                   & 0                   & 0
	\end{array}\right].
\end{equation}
Looking at the form of $E_{ij}$ we see that the same procedure for 
calculating the direction of the wave in general relativity holds for all 
the observer-independent classes: the eigenvector of $\bbox{A}$ with 
eigenvalue equal to zero points at the source.  The one exception to this 
statement is the case where the driving forces remain in a fixed line, for 
example $\Im\Psi_4=0, \Re\Psi_4 =\Phi_{22}$.  In this situation the 
direction of the wave can only be determined within the plane defined by 
the two eigenvectors with eigenvalues equal to zero.

\subsection{Detector response to impulsive excitations}
\label{sec:impulse}

Impulsive excitations are often used on resonant-mass detectors to 
calibrate the antenna \cite{Mauceli_PRD_1996}.  The excitations are usually 
administered by either a short electrical burst applied to a calibrator 
attached to the surface or a hammer blow.  Impulsive excitations were also 
used to test the analysis techniques used for experiments with the 
prototype spherical antenna at Louisiana State 
University \cite{Merkowitz_PRD_1996,Merkowitz_PRD_1997}.

A radial impulse excitation can be easily described if we choose the $z'$ 
axis to be along the direction of the impulse.  By examining the quadrupole 
eigenfunctions of the sphere in this frame we notice that out of these 
modes only the $\Psi_{125}$ mode will be excited (other sphere modes will 
also be excited but their response can be removed by narrow-band 
filtering).  All of the other quadrupole modes have a vanishing radial 
component of their eigenfunctions at this location which makes their 
``overlap'' integral with the impulse vanish.  In this frame the detector 
response is
\begin{equation}
   \bbox{A}' 
   =
   \left[\begin{array}{ccc}
      -\frac{1}{\sqrt{3}}g_{5}' & 0 &  0 \\
      0 &  -\frac{1}{\sqrt{3}}g_{5}' &  0 \\
      0 &  0 &  \frac{2}{\sqrt{3}} g_{5}'
   \end{array} \right].
   \label{eqn:impulse_matrix}
\end{equation}  
In the lab frame $\bbox{A}$ is still given by 
Eq.~(\ref{eqn:cartesian_strain_tensor}).

Again, the direction can be found by calculating the eigenvalues and 
eigenvectors of the lab frame $\bbox{A}$.  In the absence of noise, the 
eigenvector corresponding to the direction has a non-zero eigenvalue that 
is opposite in sign and twice as large as the two other eigenvalues.

\section{Solution to the inverse problem in the presence of noise}
\label{sec:noisy}

We now return to the case of a gravitational wave in general relativity to 
solve the inverse problem in the presence of noise.  At the end of this 
section we present the application of this solution to the other types of 
excitations mentioned above.  For this discussion we assume that the mode 
channels $\bbox{g}$ are independent and have normally distributed noise 
with the same variance.  This is a reasonable assumption as the truncated 
icosahedral arrangement of identical transducers ideally satisfies these 
conditions \cite{Johnson_PRL_1993}.  In addition, several other proposals 
of transducer arrangements also produce independent mode channels 
\cite{Zhou_PRD_1995,Lobo_CQG_1998} (but the sensitivity of each mode 
channel is different under normal conditions \cite{Stevenson_PRD_1997}).

\subsection{Solution for general relativity}

Noise in the mode channels $\bbox{g}$ will change the eigenvalues and 
eigenvectors of $\bbox{A}$ such that they are no longer equal to those of 
$\bbox{H}$.  To gain some insight into this situation, let us consider the 
noise as a perturbation $\bbox{N}$ to the matrix $\bbox{H}$
\begin{equation}
	\bbox{A} = \bbox{H} + \bbox{N}.
	\label{eqn:strain_and_noise}
\end{equation}
The matrix $\bbox{N}$ is constructed from the noise in each mode channel 
$\bbox{g}$, thus it has the same form as 
Eq.~(\ref{eqn:cartesian_strain_tensor}).  The matrix $\bbox{A}$ is 
therefore still symmetric and traceless and has the eigenvalue equation
\begin{equation}
	\bbox{A}\bbox{x}' = \kappa_{x'}\bbox{x}'.
	\label{eqn:noisy_ev}
\end{equation}
The eigenvectors of $\bbox{H}$ can be expanded in terms of the 
eigenvectors of $\bbox{A}$
\begin{equation}
	\bbox{x} = \sum_{x'}{C_{xx'} \bbox{x}'},
\end{equation}
where the matrix $\bbox{C}$ is close to the identity matrix if the 
perturbation is small.  However, since we do not know the values of the 
matrix $\bbox{N}$ we cannot calculate any corrections to the matrix 
elements $C_{xx'}$, thus the best approximation to $\bbox{x}$ we can find 
is $\bbox{x}'$.

Also from perturbation theory we see that the eigenvalue corresponding to 
the estimated direction of the wave is $\kappa_{x'}\approx 0$ if the 
perturbation is small.  Its magnitude will increase as the SNR decreases, 
but it should remain smaller than the other two eigenvalues of $\bbox{A}$ 
for SNR $>1$.  Consequently the eigenvector corresponding to the estimated 
direction of the wave can be selected from the three eigenvectors of 
$\bbox{A}$ by choosing the one whose eigenvalue is ``closest'' to zero.  
Once $\bbox{x}'$ is found, it can be used to estimate the direction of the 
source using Eqs.~(\ref{eqn:gamma}) and~(\ref{eqn:beta}).

The perturbation approach gives us a conceptual feel for the solution, but 
a more rigorous proof seems necessary.  The problem we wish to solve is to 
estimate the direction and polarization that makes the measured five mode 
channels $g_{m}$ most ``look like'' the expected signal from a 
gravitational wave, $h_{m}$ from Eqs.~(\ref{eqn:h_all}).  Zhou and 
Michelson used a statistical argument to justify using the least square 
error in their maximum likelihood method to fit for the direction and 
polarization \cite{Zhou_PRD_1995}.  Given the poor statistics in this 
estimation (only five samples) one might question their statistical 
approach, nevertheless the least squares error seems to be a reasonable 
choice to make under the conditions on the noise stated above.

The least squares error can be written as
\begin{equation}
	Q = \sum_{m=1}^{5}{\left(g_{m}-h_{m}\right)^{2}}.
	\label{eq:LS}
\end{equation}
The values of $h_{+}$ and $h_{\times}$ that minimize $Q$ can be found by 
simultaneously solving the equations ${\partial Q}/{\partial h_{+}} = 0$ 
and ${\partial Q}/{\partial h_{\times}} = 0$.  Doing so using 
Eqs.~(\ref{eqn:h_all}) we find
\begin{equation}
	h_+ =
	  g_{1} \frac{1}{2}(1+\cos^{2}{\beta})\cos{2\gamma} 
	- g_{2} \frac{1}{2}(1+\cos^{2}{\beta})\sin{2\gamma}
	- g_{3} \frac{1}{2}\sin{2\beta}\sin{\gamma}
	+ g_{4} \frac{1}{2}\sin{2\beta}\cos{\gamma}
	+ g_{5} \frac{\sqrt{3}}{2}\sin^{2}{\beta},
	\label{eqn:h_p_LS}
\end{equation}
\begin{equation}
	h_\times =
	g_{1} \cos{\beta}\sin{2\gamma} + 
	g_{2} \cos{\beta}\cos{2\gamma} +
	g_{3} \sin{\beta}\cos{\gamma} + 
	g_{4} \sin{\beta}\sin{\gamma}.
	\label{eqn:h_x_LS}
\end{equation}
Note that Eqs.~(\ref{eqn:h_p_LS}) and~(\ref{eqn:h_x_LS}) are identical to 
Eqs.~(\ref{eqn:h_p}) and~(\ref{eqn:h_x}) found for the noiseless case.  
This connection will be useful below.

We might also look for the minimum of $Q$ with respect to the direction of 
the wave by taking partial derivatives with respect to $\beta$ and 
$\gamma$.  This procedure leads to very complicated non-linear equations 
whose solution is not easily obtained.  For this reason we instead will 
look at how $Q$ varies close to our eigenvector solution.  We begin by 
rewriting $Q$ in terms of the detector response
\begin{equation}
	Q = \frac{1}{2}\sum_{j=1}^{3}\sum_{i=1}^{3}{\left(A_{ij} - h_{ij}\right)^2}
\end{equation}
\begin{equation}
	  = \frac{1}{2}\text{Tr}\left(\left[\bbox{A}-\bbox{H}\right]
	    \left[\bbox{A}-\bbox{H}\right]^{T}\right)
\end{equation}
\begin{equation}
	  \equiv 
	  \frac{1}{2}\|\bbox{A}-\bbox{H}\|^{2}.
\end{equation}
The inner product $\|\bbox{A}-\bbox{H}\|^{2}$ can be interpreted as the 
distance between $\bbox{A}$ and $\bbox{H}$ which we know to be invariant to 
rotations.  If $\bbox{R}$ is the matrix that diagonalizes $\bbox{H}$ such 
that $\bbox{H}' = \bbox{R}^{T}\bbox{H}\bbox{R}$, we can write
\begin{equation}
	Q = \frac{1}{2}\|\bbox{R}^{T}\bbox{A}\bbox{R}-\bbox{H}'\|^{2}.	
  \label{eq:LS_diag}
\end{equation}
It is now clear that the least squares fit is the matrix $\bbox{R}$ that 
minimizes the distance $Q$.  Geometrically, this minimum occurs when 
$\bbox{A}'=\bbox{R}^{T}\bbox{A}\bbox{R}$ is the projection of $\bbox{H}'$ 
onto $\bbox{A}$.  Given that $\bbox{H}'$ is diagonal one might guess that 
this minimum occurs when $\bbox{A}'$ is also diagonal.  Let us proceed to 
prove this conjecture.

Let $\bbox{R}_{0}$ be the matrix constructed from the eigenvectors of 
$\bbox{A}$ so that $\bbox{R}_{0}^{T}\bbox{A}\bbox{R}_{0}=\bbox{D}$ where 
$\bbox{D}$ is a diagonal matrix.  Let us also assume $\bbox{R}_{0}$ 
differs from $\bbox{R}$ by a small rotation $\epsilon\bbox{W}$ such that
\begin{equation}
	\bbox{R} = 
	\bbox{R}_{0}
	[\bbox{I}+\epsilon\bbox{W}+\frac{1}{2}\epsilon^{2}\bbox{W}^{2}],
	\label{eq:R}
\end{equation}
where $\bbox{W}$ is a skew-symmetric matrix with zeros along the diagonal.  
Substituting Eq.~(\ref{eq:R}) into Eq.~(\ref{eq:LS_diag}) and keeping terms 
only up to $\epsilon^{2}$ we find
\begin{equation}
	Q \simeq 
	\frac{1}{2}\|
	\bbox{D}- \bbox{H}'
	+ \epsilon[\bbox{D}\bbox{W}-\bbox{W}\bbox{D}]
	+ \frac{1}{2}\epsilon^{2}
	[ \bbox{D}\bbox{W}^{2}
	-2\bbox{W}\bbox{D}\bbox{W}
	+ \bbox{W}^{2}\bbox{D}]
	\|^{2},
	\label{eq:LS_e2}
\end{equation}
Expanding Eq.~(\ref{eq:LS_e2}) and remembering that the trace is invariant 
under cyclic permutations of the matrices in a product and that 
$\bbox{H}'\bbox{D}=\bbox{D}\bbox{H}'$ we find
\begin{equation}
	Q \simeq
	\frac{1}{2}\|\bbox{D}- \bbox{H}'\|^{2} 
	+\epsilon^{2}\text{Tr}(
	 \bbox{H}'\bbox{W}\bbox{D}\bbox{W}	
	-\bbox{D}\bbox{H}'\bbox{W}^2
	).
\end{equation}
All the first order terms in $\epsilon$ have vanished so we have proven 
that $Q$ is stationary near $\bbox{R}=\bbox{R}_{0}$.  To show this point is 
a minimum we need to evaluate the second order terms in $\epsilon$.

$\bbox{W}$ can be written in terms of a unit vector $\bbox{n}$ representing 
the axis of rotation, so the square of this matrix is given by
\begin{equation}
	[\bbox{W}^{2}]_{ij} = n_{i}n_{j} - \delta_{ij}.
	\label{eq:W2}
\end{equation}
Recalling the procedure for deriving Eqs.~(\ref{eqn:h_p}) 
and~(\ref{eqn:h_x}) in the noiseless case and that they are identical to the 
least squares minimum Eqs.~(\ref{eqn:h_p_LS}) and~(\ref{eqn:h_x_LS}) we can 
set 
\begin{equation}
	H_{11}'=-H_{22}'=\frac{1}{2}(D_{11}-D_{22}).
\end{equation}
The matrix $\bbox{D}$ is also traceless so $D_{11}+D_{22}+D_{33}=0$.  Now 
the second order terms can be written as
\begin{equation}
	O(\epsilon^2) 
	=
	\frac{1}{2}\left(
	 D_{11}^2(2-3n_1^2)
	+D_{22}^2(2-3n_2^2)
	+D_{11}D_{22}(-4+3n_1^2+3n_2^2)
	\right)
	\label{eq:Oe2}
\end{equation}
Using $n_1^2 + n_2^2 + n_3^2 = 1$ and $0 \leq n_i^2 \leq 1$ we find that 
$O(\epsilon^2)$ is not guaranteed to be positive for all possible real 
values of $D_{11}$ and $D_{22}$.  Fortunately we may also assume we have 
ordered the eigenvalues of $\bbox{A}$ such that $D_{33}$ is the eigenvalue 
closest to zero.  Now we have an additional condition $D_{11} = -bD_{22}$, 
where $0.5<b<2$.  Substituting this into Eq.~(\ref{eq:Oe2}) we find
\begin{equation}
	O(\epsilon^2) 
	=
	\frac{1}{2}
	D_{22}^2
	\left(b+1\right)
	\left(2b + 2 - 3bn_1^2 - 3n_2^2\right).
\end{equation}
By inspection we see that $O(\epsilon^2)$ is always positive under the 
conditions stated above, therefore $Q$ is always a minimum near 
$\bbox{R}=\bbox{R}_{0}$.

We further used a Monte Carlo type simulation to show that this point is 
always the global minimum of $Q$.  For a wave of a given direction and 
polarization we calculated the spherical amplitudes $h_{m}$ and added a 
random number (variance $\sigma_h^2$ and zero mean) to obtain the mode 
channels.  The direction and polarization were estimated using the 
eigenvector method as well as by numerically finding the minimum of $Q$ 
from Eq.~(\ref{eq:LS}).  We found the two methods gave identical results, 
even for high values of $\sigma_h^2$, confirming that 
$\bbox{R}=\bbox{R}_{0}$ is a global minimum of $Q$.  Therefore, the 
diagonal form of $\bbox{A}$ is the best approximation to $\bbox{H}'$ and 
can be used to estimate the direction and polarization of the wave.

Both the maximum likelihood method and the eigenvector solution minimize 
the mean square error under the conditions on the noise stated above, 
therefore, produce the same answer for the estimated values.  However, the 
eigenvector solution is more straightforward and computationally simple.  
We construct the matrix $\bbox{A}$ from the mode channels $\bbox{g}$ and 
compute its eigenvalues $\kappa_{x'}$ and eigenvectors $\bbox{x}'$.  We 
choose the eigenvector with eigenvalue closest to zero and estimate the 
direction of the wave using Eqs.~(\ref{eqn:gamma}) and~(\ref{eqn:beta}).  
The polarization amplitudes can be estimated using Eqs.~(\ref{eqn:h_p_LS}) 
and~(\ref{eqn:h_x_LS}).

\subsection{Extensions of the eigenvector solution}

The detector response matrix $\bbox{A}$ for other metric theories of 
gravity as well as for impulse excitations satisfy the symmetry arguments 
used in the discussion for general relativity.  This means we can easily 
adapt the noiseless solutions to the case where noise is present in the same 
fashion.

For observer-independent gravitational theories the method for estimating 
the direction of the wave is identical to that of general relativity: the 
eigenvector of the detector response matrix with eigenvalue closest to zero 
can be used to estimate the direction of the source.  Observer-dependent 
theories require prior knowledge of the polarization states of the wave 
before any estimate of the direction can be made.  Once these are known it 
should be straightforward to adapt the eigenvector technique to estimate 
the direction.  In the case of an impulsive excitation, the eigenvector 
corresponding to the direction has an eigenvalue that is opposite in sign 
and greater in magnitude than the two other eigenvalues.  Converting the 
eigenvectors to a direction again comes from Eqs.~(\ref{eqn:gamma}) 
and~(\ref{eqn:beta}).

\section{Discussion}
\label{sec:discussion}

The eigenvector solution is very convenient in that the inverse problem is 
reduced to solving a trivial eigenvalue problem.  The solution is 
computationally simple, making this technique very efficient for use in an 
automated data analysis system.  This feature may be important if one 
considers using a large number of candidate gravitational wave events in a 
coincidence exchange between several detectors where the source direction 
is used as a criterion to veto excess coincidences.

The main restrictions on the eigenvector solution is that the mode channels 
must be independent and the noise normally distributed with equal variance.  
These restrictions can ideally be satisfied for a number of transducer 
arrangements~\cite{Johnson_PRL_1993,Zhou_PRD_1995}.  We found that the 
eigenvector solution corresponds exactly to the maximum likelihood method 
under these conditions.  It may be possible to apply this solution when the 
noise is not gaussian or is different for each mode channel, however 
further research is necessary to verify this extension.

\begin{figure}
\centerline{\psfig{file=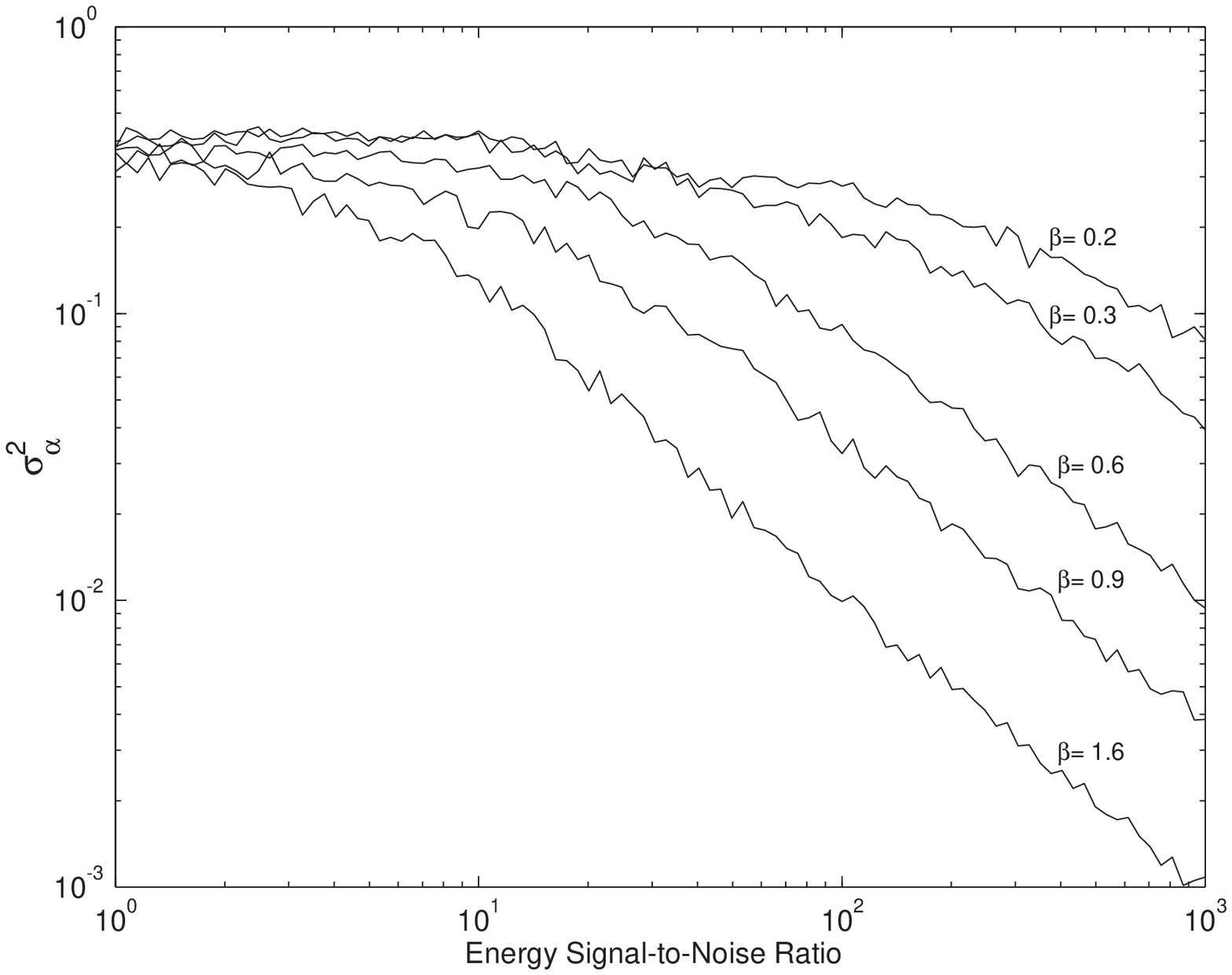,width=3.0in}}

\caption{The results of a numerical simulation describing the variance of 
the polarization angle $\alpha$ for a range of SNR and several values of 
the direction angle $\beta$.  Each line was computed by a 500 trial Monte 
Carlo simulation for 100 logarithmically spaced SNR for the corresponding 
value of $\beta$.}

\label{fig:snr_alpha}
\end{figure}

Through a number of numerical simulations as well as examination of the 
work of others \cite{Zhou_PRD_1995,Stevenson_PRD_1997} we found that the 
errors due to the noise on a direction estimation are independent of the 
source location and wave amplitude for a given SNR.  However, the 
estimation of the polarization amplitudes using Eqs.~(\ref{eqn:h_p_LS}) 
and~(\ref{eqn:h_x_LS}) lead to direction dependent uncertainties.  For 
example, Fig.~\ref{fig:snr_alpha} shows the variance on the polarization 
angle $\alpha = \tan^{-1}(h_\times/h_+)$ for a range of SNR and several 
values of $\beta$ found from a Monte Carlo type simulation.  Notice that 
the variance increases for low values of $\beta$.  This realization is 
disturbing given that a spherical antenna is equally sensitive to waves 
from all directions and polarizations.

One might consider using a different coordinate system to try to avoid the 
directions with very poor estimates of the two polarization amplitudes.  
For example, use the xyz-convention of the Euler angles where the first and 
the last rotations are not the same.  This actually will not solve our 
problem, but instead change the directions in the sky which lead to the 
poor estimates.  If we transform back from this coordinate system to the 
y-convention we just reintroduce the errors and thus have gained nothing.

This dependency on the source direction is not unique to a sphere, a 
network of bars or interferometers will also suffer from this problem 
\cite{Gursel_PRD_1989}.  This leads us to believe that we are excluding a 
piece of information from our procedures.  The solution may lie in using 
the information from the two other eigenvectors of the detector response.  
In the above derivations these eigenvectors were simply discarded, but they 
also contain information about the gravitational wave that may eliminate 
these direction dependent errors.  This approach will be the topic of a 
future paper~\cite{Lobo_PRD_1999}.

While there are a few limitations to the eigenvector solution, its 
simplicity makes it easily extendable to other types of excitations.  As 
discussed above, impulsive excitations can be located using this technique.  
As a practical example we recall that this solution was successfully tested 
on experiments with the LSU prototype spherical antenna 
\cite{Merkowitz_PRD_1997}.  This practical confirmation of its validity 
gives us the confidence that it can be implemented on a real spherical 
antenna searching for gravitational waves.

\acknowledgments

I thank M.~Bassan, M.~Bianchi, E.~Coccia, and E.~Mauceli for many useful 
discussions on this work.  In particular, I thank J.~A.~Lobo for his 
assistance and advice.  I also thank the referee for suggesting part of the 
argument presented in Sec.~\ref{sec:noisy}.


\begin{table}
\begin{tabular}{lll}
Class   & Allowable polarization states    & Example \\
\hline
$II_6$  & $\Phi_{22}$, $\Psi_4$, $\Psi_3$, $\Psi_2$ & Most general \\
$III_5$ & $\Phi_{22}$, $\Psi_4$, $\Psi_3$           & Kaluza-Klein \\
$N_3$   & $\Phi_{22}$, $\Psi_4$                     & Brans-Dicke \\
$N_2$   & $\Psi_4$                                  & General relativity \\
$O_1$   & $\Phi_{22}$                               & Purely scalar \\
$O_0$   & None                                      & No wave \\
\end{tabular}
\caption{The E(2) classification scheme.}
\label{tab:E2_class}
\end{table}

\end{document}